# Conduction band tuning by controlled alloying of Fe into $Cs_2AgBiBr_6$ double perovskite powders

*Huygen J. Jöbsis[1], Kostas Fykouras[2], Joost W.C. Reinders[1], Jacco van Katwijk[1], Joren M. Dorresteijn[1], Tjom Arens[1], Ina Vollmer[1], Loreta A. Muscarella[1], Linn Leppert[2]\*\*, and Eline M. Hutter[1]\**

1. Inorganic Chemistry and Catalysis, Department of Chemistry, Utrecht University, Princetonlaan 8, 3584 CB Utrecht, the Netherlands.
2. MESA+ Institute for Nanotechnology, University of Twente, 7500 AE Enschede, The Netherlands.

Authors to whom correspondence should be addressed:

Eline M. Hutter, *e.m.hutter@uu.nl

Linn Leppert, **l.leppert@utwente.nl

Halide double perovskite semiconductors such as $Cs_2AgBiBr_6$ are widely investigated as a more stable, less toxic alternative to lead-halide perovskites in light conversion applications including photovoltaics and photoredox catalysis. However, the relatively large and indirect bandgap of $Cs_2AgBiBr_6$ limits efficient sunlight absorption. Here, we show that controlled replacement of $Bi^{3+}$ with $Fe^{3+}$ via mechanochemical synthesis results in a remarkable tunable absorption onset between 2.1 and ~1 eV. Our first-principles density functional theory (DFT) calculations suggest that this bandgap reduction originates primarily from a lowering of the conduction band upon introduction of $Fe^{3+}$. Furthermore, we find that the tunability of the conduction band energy is reflected in the photoredox activity of these semiconductors. Finally, our DFT calculations predict a direct bandgap when >50% of $Bi^{3+}$ is replaced with $Fe^{3+}$. Our findings open new avenues for enhancing the sunlight absorption of double perovskite semiconductors and for harnessing their full potential in sustainable energy applications.

## 1. Introduction

Over the past years halide double perovskites, also known as elpasolites, have gained increasing attention as less toxic, more stable semiconductors than the well-studied lead-halide perovskites.[1–6] The term double perovskite is often used because its crystal structure, $A_2M^IM^{III}X_6$, resembles a perovskite crystal structure in which, the divalent cation is replaced by alternating monovalent ($M^I$) and trivalent ($M^{III}$) cations. In particular, $Cs_2AgBiBr_6$ (CABB) has been widely studied for radiation (*e.g.* X-ray) detectors, due to its excellent stability and good absorptivity of high energy photons.[7] Its application in single junction outdoor photovoltaics remains challenging, due to the relatively large indirect bandgap (2.0 – 2.3 eV), and rapid localization of mobile charges.[8–11] However, recent work reports CABB-based solar cell devices with photoconversion efficiencies of 6.37% under AM 1.5 illumination[12] and first-principles studies predict the material to have a spectroscopic limited maximum efficiency of 10.5%.[13] Although this is still far from competing with lead-halide perovskites, it is comparable to the performance of other lead-free perovskite devices, such as tin-halide perovskite solar cells.[5] In addition, the relatively large bandgap can be beneficial for indoor photovoltaic applications, because of the better match of the double perovskite absorption spectrum and the emission spectrum of *e.g.* white LEDs.[14,15] Furthermore, CABB was shown to be a suitable photoactive material for photoredox chemistry, including $CO_2$ reduction and $H_2$ evolution reactions.[16–18] The photoconversion efficiency for both photovoltaic and photoredox applications is, however, limited by the relatively poor absorption over the indirect bandgap. The search for non-toxic, stable double perovskites with direct bandgaps in the visible will therefore be of major interest for several photoconversion applications. Finally, finding routes to control the bandgap are desirable both for multi-junction photovoltaics and photoredox chemistry. For the latter, tuning the absolute energies of the valence and conduction band may be used as a strategy to match the electron and hole energies with the chemical potential of the corresponding reduction and oxidation reactions.

Similar to lead-halide perovskites, the bandgap of double perovskites can be manipulated through (partial) substitution of one (or more) ion(s) with a similarly charged ion.[19] In previous reports it was demonstrated that the bandgap energy of CABB can be increased upon partly replacing $Bi^{3+}$ with $In^{3+}$ or decreased using $Sb^{3+}$ as a $Bi^{3+}$-substituent.[20–23] Moreover, it was shown that substitution of $Bi^{3+}$ with other metal cations affects the nature of the bandgap.[24–26] For example, increasing of the $In^{3+}$-content in CABB creates direct transitions in the material, which is in principle favorable for light absorption and emission. However, the enlargement of the bandgap by $In^{3+}$-alloying leads to an overall reduction of visible light absorption.

Recent work has demonstrated that the (partial) replacement of $Bi^{3+}$ or $In^{3+}$ with $Fe^{3+}$ extends the absorption of visible light into the near infrared,[27,28] which can potentially improve the photoconversion efficiencies of these materials. Moreover, the introduction of $Fe^{3+}$ gives rise to a temperature-dependent magnetic response, rendering these materials interesting for spintronic devices.[28,29] Despite these promising material properties, the number of reports in literature remains scarce. This might be due to the relatively difficult synthesis of these materials via solvent-chemistry approaches, requiring a common solvent for four different precursor salts to avoid the crystallization of undesired side phases.[6,20,30] In addition, previous routes for incorporating $Fe^{3+}$ in CABB have led to the formation of Fe-rich clusters, rather than homogeneous alloying with $Bi^{3+}$.[28]

Recent work has shown that solvent-free, mechanochemical synthesis is a powerful method to obtain alloyed elpasolites with high control over composition.[31,32] Here we demonstrate that such a mechanochemistry approach can be used to synthesize $Cs_2AgBi_{1-x}Fe_xBr_6$ with different Bi:Fe ratios. With the use of X-ray diffraction and elemental analysis, we find that this synthesis method leads to an incorporation of up to 30% $Fe^{3+}$ (i.e. $x = 0.3$) with homogeneous distribution, while preserving the elpasolite crystal structure. In addition, we observe that substituting $Bi^{3+}$ with $Fe^{3+}$ gradually red-shifts the absorption onset, leading to full bandgap tunability between ~2 eV (CABB) and ~1 eV ($Cs_2AgBi_{0.7}Fe_{0.3}Br_6$). To assess the effect of the bandgap changes on the material's performance in photoredox reactions, we performed methylene blue dye decolorization experiments. For all compositions, we observed significant depletion of dye in solution. From dark control experiments, we find that only $Cs_2AgBiBr_6$ is photocatalytically active, and degrades methylene blue via the formation of superoxide from oxygen. For $Cs_2AgBi_{1-x}Fe_xBr_6$, the dye depletion mainly originates from adsorption of the dye on the double perovskite surface rather than degradation of the dye. These experimental findings are explained by first-principles density functional theory (DFT) calculations of the bandstructure of the alloyed materials. These calculations demonstrate that the observed bandgap reduction is due to lowering of the conduction band minimum by Fe 3d states, while the valence band maximum remains constant. Furthermore, our calculations predict that $Cs_2AgBi_{1-x}Fe_xBr_6$ changes from an indirect to a direct bandgap material between $x$ = 0.5 and $x$ = 0.75. At higher $Fe^{3+}$ concentrations, hybridization between conduction band states derived from Ag 5s and Fe 3d result in a disperse conduction band minimum, highlighting the potential of these materials as (sun-)light absorbers, which encourages future research to make these materials experimentally. Hence, our DFT calculations suggest that the lack of photocatalytic performance is due to the conduction band energy being too low to react with oxygen. However, the combination of strong surface adsorption, visible light absorption, and magnetic character warrants further investigation of $Cs_2AgBi_{1-x}Fe_xBr_6$ for catalytic and optoelectronic applications.

## 2. Results and Discussion

To synthesize $Cs_2AgBi_{1-x}Fe_xBr_6$ powders we use a ball mill equipped with stainless steel jars.[33] The ball mill jar is filled with stainless steel beads (10 mm diameter) and the different reactants (CsBr, AgBr, $BiBr_3$ and $FeBr_3$) in stoichiometric ratios. During this solvent-free milling process, the reactants are finely ground, increasing the reaction surface area. The combination of the large force and increased surface area that is created inside the ball mill allows to surpass (side-)phases that are energetically more favorable at atmospheric conditions.[34] By using different $BiBr_3$:$FeBr_3$ ratios we were able to synthesize various $Cs_2AgBi_{1-x}Fe_xBr_6$ compositions with $x$ = 0, 0.01, 0.1, 0.15 and 0.3. The Bi:Fe ratio in the resulting powders were determined using scanning electron microscopy energy dispersive X-ray spectroscopy (SEM–EDX) showing a good agreement between the expected ratios based on the reactants used (Figure S1). For higher $Fe^{3+}$ concentrations ($x \geq 0.3$) no phase-pure double perovskites were obtained as will be discussed below. A detailed description of the mechanochemical synthesis method is provided in the Supplementary Information.

The formation of the double perovskite crystal structure is confirmed using powder X-ray diffraction (XRD) crystallography (Figure 1a). Introduction of the smaller $Fe^{3+}$ cations (ionic radius of 0.65 Å compared to the ionic radius of $Bi^{3+}$ of 1.17 Å)[35] into the cubic crystal structure shifts the diffraction pattern to higher diffraction angles indicating a reduction of the lattice parameter, $a$ (Figure 1b). For all compositions, $a$ is determined using profile refinement considering only the double perovskite phase ($Fm\bar{3}m$ symmetry) (Figure S2 and Table S1).[3] The linear reduction of $a$ as a function of the Bi:Fe ratio indicates that during the ball mill synthesis solid solutions are formed (see red points in Figure 1c). If $Fe^{3+}$ is indeed homogeneously incorporated at the $Bi^{3+}$ position, the reflection intensity of, e.g. the (200) plane (comprising $Ag^+$, $Bi^{3+}$ and $Br^-$), should decrease faster with respect to the (111) plane (comprising just $Ag^+$) as the scattering factor of $Bi^{3+}$ is larger than for $Fe^{3+}$ (Figure 1d).[36] To confirm this, the theoretical diffractograms of $Cs_2AgBi_{1-x}Fe_xBr_6$ (for various values of $x$) were calculated using VESTA (Figure S3). Figure 1c shows a good agreement between the ratio of reflection intensities of the (222) and (400) planes for the experimental and calculated diffractograms. Hence, we conclude $Cs_2AgBi_{1-x}Fe_xBr_6$ solid solutions can be formed up to $x$ = 0.3 using ball milling. For $x$ > 0.3, a deviation from this trend is observed which is possibly due to incomplete incorporation of $Fe^{3+}$ or local compositional variations. According to an adapted version of the tolerance factor the fully exchanged $Cs_2AgFeBr_6$ might be a stable double perovskite.[37] However, thermodynamically this composition is predicted to be unstable.[38] The stability of halide perovskites is primarily determined by the A- and X-site elements, following the general trend that the stability increases when X becomes lighter. Consequently, in literature stable compositions of the fully substituted $Cs_2AgFeCl_6$ are reported[39–41] whereas in this work, using bromide, no phase-pure compositions were observed for $x$ > 0.3 (Figure S4).

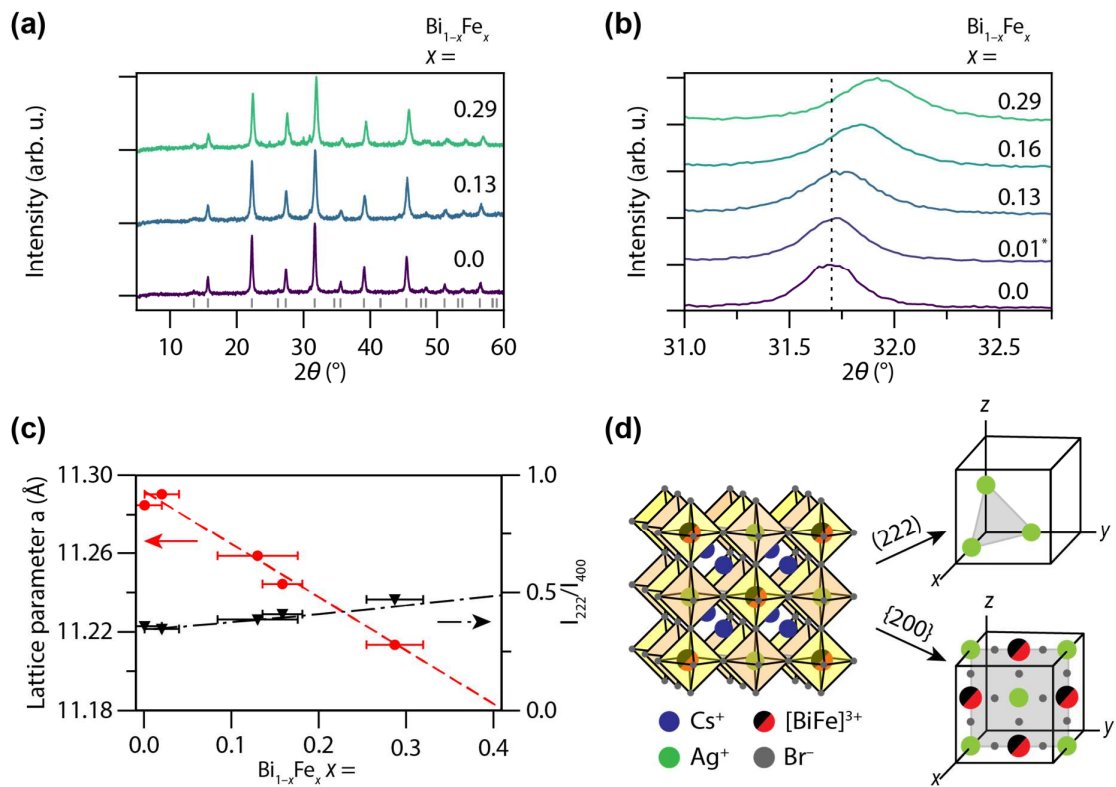

**Figure 1. Structural properties of $Cs_2AgBi_{1-x}Fe_xBr_6$. a)** Powder XRD of $Cs_2AgBi_{1-x}Fe_xBr_6$ for $x$ = 0.0, 0.09 and 0.29 (colored). The grey dashes indicate the positions of the calculated reflections for the $Cs_2AgBiBr_6$ (CABB) crystal structure. **b)** Zoom in of **a)** showing a shift of the diffractograms upon increasing the $Fe^{3+}$-content. **c)** Reduction of the lattice parameter, $a$, (red dots) as a function of the Bi:Fe ratio (left axis). On the right axis the ratio of reflection intensity off the (222) and (400) planes are plotted as a function of Bi:Fe ratio for the experimental data (black triangles) and calculated diffractograms (dash–dotted black line). **d)** Graphical representation of the double perovskite crystal structure, showing the (222) and (200) planes. The latter is an equivalent of (400) hence the {200} notation. *For $x$ = 0.01 the concentration of $Fe^{3+}$ was below the detection limit of the SEM-EDX.

From Figure 2a it is apparent that the visible light absorption increases with $Fe^{3+}$ incorporation as the color of CABB changes from orange to brown to black. Figure 2b presents the Kubelka–Munk transform of the diffuse reflectance spectra of these powder materials. The resulting approximations of the absorption coefficient profile show that the absorption onset is red-shifted by 1 eV when going from $x$ = 0 to $x$ = 0.3. The broad feature of the absorption profile towards the NIR, however, hints at a change in the nature of the absorption onset. For CABB, broad photoluminescence (PL) is typically observed several hundreds of meV below the bandgap (Figure 2c). Upon introduction of $Fe^{3+}$, we observe a strong decrease in PL intensity which is in line with previous observations in the literature.[27] For $x \leq$ 0.01 the PL spectrum is slightly red-shifted, however, to a smaller extent than expected based on the red-shift observed in the absorption spectrum. This suggests that the recombination mechanism of CABB remains unchanged on the addition of trace amounts of $Fe^{3+}$. For $x \geq 0.1$ on the other hand the PL is quenched in such a way that the spectra are dominated by artifacts of the equipment (Figure S5). The loss of PL upon the introduction of $Fe^{3+}$ suggests the formation of additional non-radiative recombination pathways.

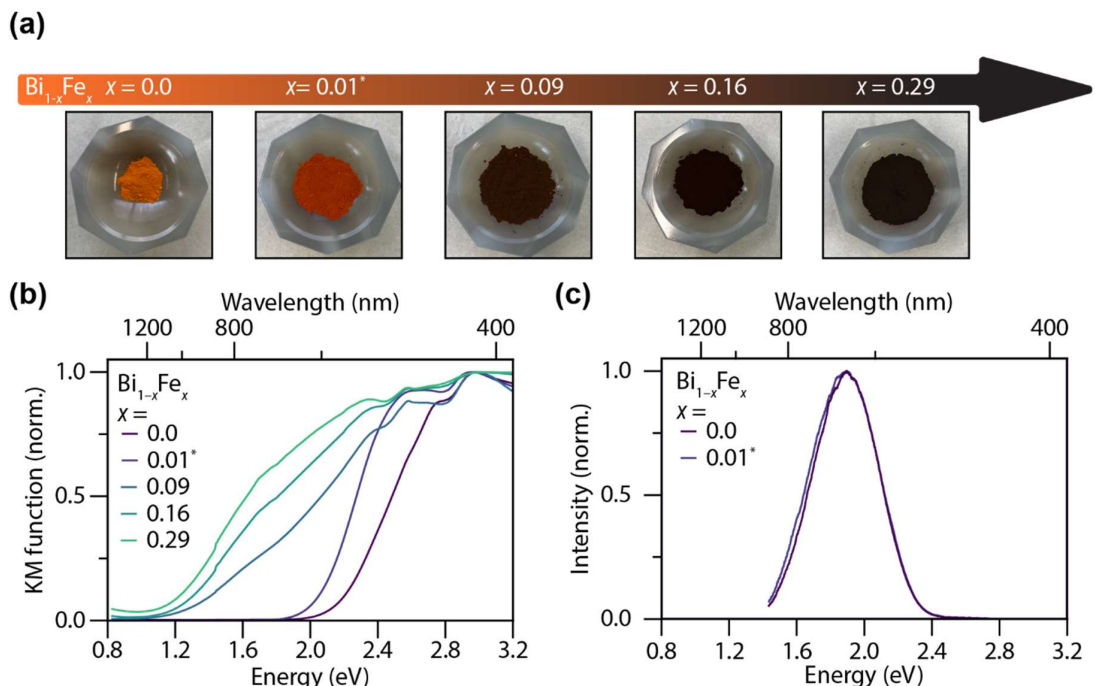

**Figure 2. Optical properties of $Cs_2AgBi_{1-x}Fe_xBr_6$. a)** Incorporation of $Fe^{3+}$ into CABB strongly red-shifts the absorption of visible light. **b)** Kubelka–Munk transform of the diffuse reflectance spectra recorded on the $Cs_2AgBi_{1-x}Fe_xBr_6$ samples with ($x$ = 0.0, 0.01, 0.09, 0.16 and 0.29), showing a gradual red-shift of the absorption onset with increasing $Fe^{3+}$-content. The jump in the data around 1.4 eV is an artifact of the equipment. **c)** Photoluminescence spectra of $Cs_2AgBiBr_6$ and $Cs_2AgBi_{0.99}Fe_{0.01}Br_6$. Further increasing the $Fe^{3+}$-content quenches the PL almost completely (Figure S5). *For $x$ = 0.01 the concentration of $Fe^{3+}$ was below the detection limit of the SEM-EDX.

In literature dye decolorization (or degradation) experiments are often used to test the activity of photoactive materials.[16,42–44] Here, we perform methylene blue (MB) decolorization experiments using different light sources. Figures 3a–d show the MB concentration over time, when mixed with $Cs_2AgBi_{1-x}Fe_xBr_6$ in ethanol, in dark (black triangle) and while illuminated with green (550 nm, FWHM = 30 nm, green dot) or blue light (445 nm, FWHM = 30 nm, blue square). Here, the light is switched on after 60 minutes of mixing and stirring, indicated as t = 0. We note that the double perovskites are stable under the experimental conditions, since the XRD patterns before and after the experiment show the double perovskite crystal structure is preserved (Figure S6). For CABB, when illuminating with blue light the decolorization is completed within 150 minutes following a linear dependence on the irradiation time. The decolorization of MB may occur via the formation of superoxide radicals ($^\bullet O_2^-$), hydroxyl radicals ($^\bullet OH$) and/or direct reduction of the dye.[44] Here, we can exclude the formation of $^\bullet OH$ as the potential energy of the valence band maximum is insufficient (Figure 3e). The direct reduction of MB, i.e. direct transfer of a CABB conduction band electron to MB, is also ruled out as no activity is observed when performing the decolorization experiment under an inert atmosphere (Figure S7). Interestingly, the decolorization is significantly slower when illuminating with green or no light, confirming that this process is primarily induced by charge carriers generated in CABB, because the green light is poorly absorbed by CABB.

For the $Cs_2AgBi_{1-x}Fe_xBr_6$ compositions (Figures 3b–d), we did not observe a difference between dark and light, suggesting that the dye depletion from the solvent is dominated by dye adsorption on the double perovskite surface rather than degradation. Interestingly, the $x$ = 0.13 shows a strong interaction with the MB, adsorbing 80% of the dye within two hours, while the dye barely adsorbs onto the $x = 0.3$. The loss of photoredox activity might arise due to the lowering of the bandgap energy and/or introduction of defect states upon $Fe^{3+}$ introduction.

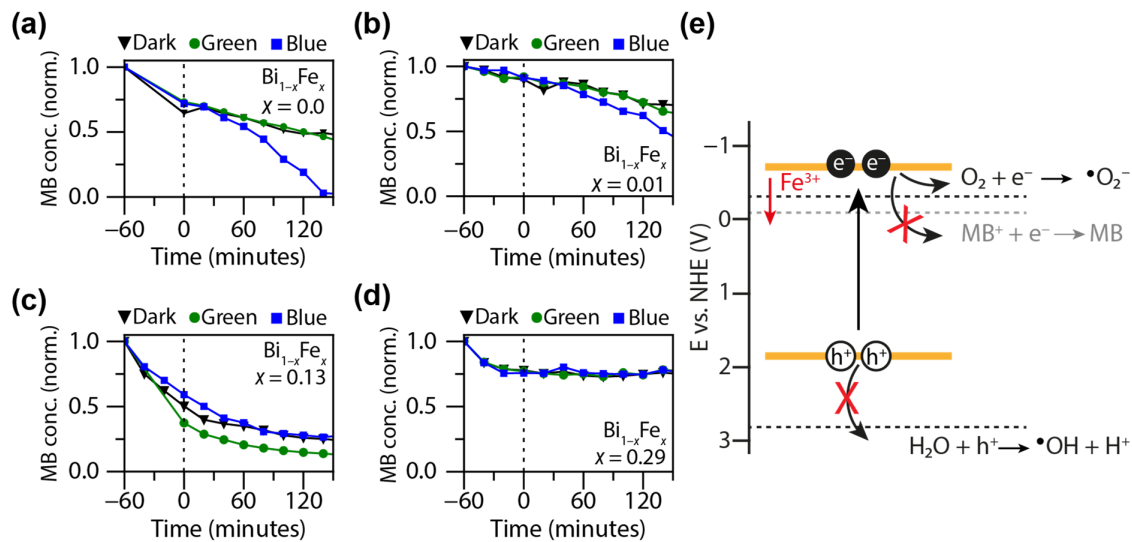

**Figure 3. Photoredox activity of $Cs_2AgBi_{1-x}Fe_xBr_6$.** Methylene blue concentration as a function of time when mixed with **a)** CABB and **b)** $Cs_2AgBi_{0.99}Fe_{0.01}Br_6$ while illuminated with blue (blue, square), green (green, dot) and no light (black, triangle). Methylene blue concentration as a function of time when mixed with **c)** $Cs_2AgBi_{0.87}Fe_{0.13}Br_6$ and **d)** $Cs_2AgBi_{0.71}Fe_{0.29}Br_6$ while illuminated with blue (blue, square), green (green, dot) and no light (black, triangle). Substitution of $Bi^{3+}$ with $Fe^{3+}$ inhibits the photochemical activity, suggesting the formation of additional defect states or the lowering of the conduction band minimum. **e)** Graphical representation of the mechanism of the decolorization of MB.

To understand how the valence and conduction band are affected by introduction of $Fe^{3+}$, we performed first-principles DFT calculations of the bandstructure of $Cs_2AgBi_{1-x}Fe_xBr_6$ for $x = 0, 0.25, 0.50, 0.75, 1$. To account for the localized nature of the Fe 3d states and correct the bandgap underestimation of Kohn-Sham DFT, we performed all bandstructure calculations with the PBE0 hybrid functional.[45] We include spin-orbit coupling (SOC) self-consistently for all compositions excluding $x = 1$, for which it has a negligible effect. All compositions are represented in a conventional 40-atom unit cell with $Fm\bar{3}m$ symmetry and fully geometry-optimized using the SCAN functional.[46] These settings result in lattice parameters in very good agreement with experiment (Figure S8), exhibiting a linear decrease of lattice parameters with increasing $x$. For $x = 0.25$ and $x = 1$, we tested the effect of the ordering of the $Fe^{3+}$ local magnetic moments on the total energies, bandstructures and bandgaps (see Figures S9.1–S9.3) . Our results are in line with a comprehensive study of magnetic order/disorder effects on the electronic structure of the closely related double perovskite $Cs_2AgFeCl_6$ by Klarbring et al.[47]

The bandstructures of all compositions are shown in Figure 4 (density of states plots can be found in Figure S9.4). In the conventional unit-cell setting, CABB ($x = 0$) has an indirect bandgap with the valence band maximum (VBM) at Γ and the conduction band minimum (CBM) at L. Our calculated bandgap is 2.48 eV with PBE0+SOC, slightly overestimating the range of experimentally reported bandgaps. Replacing one in four $Bi^{3+}$ ions for $Fe^{3+}$ ($x = 0.25$) leads to significant changes in the conduction band, while the valence band is mostly unaffected. For $x = 0.25$, we find that the CBM moves to X and is derived from Fe $3d_{z2}$ and $d_{x2-y2}$ states. From $x = 0.25$ to $x = 0.75$, we observe a gradual shift of the CBM to Γ which is due to increasing hybridization of Fe $3d_{z2}$, $d_{x2-y2}$ and Ag 5s states. At higher $x$, a pronounced change in the valence bands is due to the decreasing hybridization of Bi 6s- and Ag 4d-derived states. For $x = 1$, the VBM remains at Γ but with significantly reduced dispersion as compared to $x = 0$. Across all compositions, the Fe 3d states form a relatively flat band across the Brillouin zone, which may explain the broad absorption tail observed in our UV-vis absorption experiments as well as the quenching of the PL.

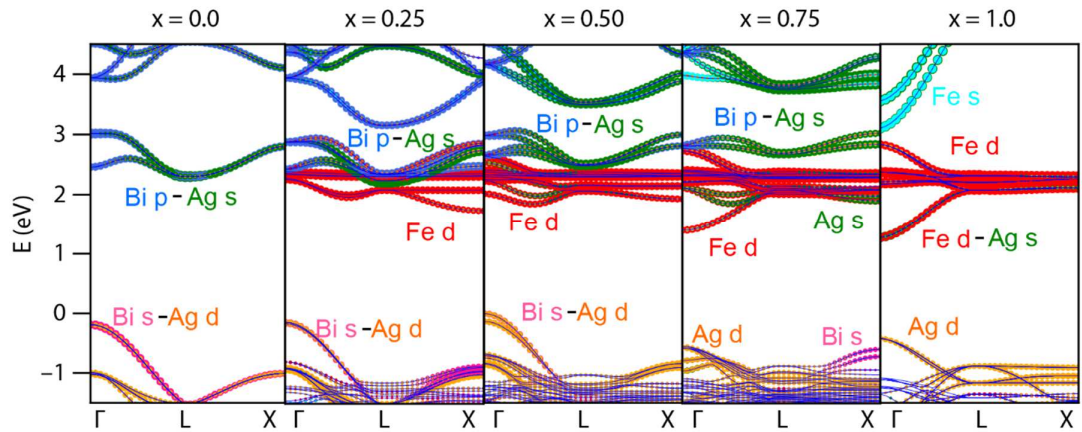

**Figure 4. PBE0+SOC bandstructures of $Cs_2AgBi_{1-x}Fe_xBr_6$ ($x = 0, 0.25, 0.50, 0.75, 1$).** Orbital contributions are indicated in color, omitting contributions of Br p for clarity. All bandstructures are aligned with respect to a low energy Cs-s state calculated at the *Γ* point. Note that all bandstructures are calculated within a conventional unit cell setting and assuming ferromagnetic spin ordering (see SI for a detailed discussion). Due to bandfolding, the VBM of CABB ($x = 0$) therefore appears at *Γ*.

We find a pronounced bandgap reduction of ~0.6 eV when $x$ is progressively increased from $x = 0$ to $x = 0.5$ (Figure S9.4). This is lower than the decrease of ~1 eV that we observe experimentally, indicating that additional defect states or excitonic effects, which are not included in our calculations, might play a role. For $x = 0.25$, we found that the ordering of local Fe magnetic moments only had a small effect on the order of 0.1 eV on the bandgap (see SI for details). Regarding the role of excitons, we note that the absorption onset of CABB was shown to be dominated by a resonant direct exciton with reported binding energies ranging between 70 – 270 meV.[11,48–51] Our bandstructure calculations allow us to predict the character of excitons in the mixed compositions based on the orbital-character of the valence and conduction band edges.[52] The disperse and isotropic CBM and VBM of the $x = 1$ material, indicate an exciton well-described by the Wannier–Mott model with a relatively low exciton binding energy of 59 meV (see SI for computational details). However, for intermediate compositions for which the lowest direct transition is between a disperse valence band and a flat conduction band at Γ, the exciton binding energy is likely larger and not well-described by the Wannier–Mott model.

The lowering of the conduction band could explain the absence of light-induced MB decolorization due to insufficient potential energy of the photoexcited electrons to form $^\bullet O_2^-$. From Figure 3e we can deduce that a reduction of the CBM of circa 300 mV will reduce the potential energy of the CBM electrons below the potential of $^\bullet O_2^-$ formation. This is in line with our DFT calculations and the red-shift of the absorption onset discussed above. In addition, it is possible that the $Fe^{3+}$ containing compositions have a higher defect density, which is in line with their lower PL intensity. The predicted low, direct bandgap, together with decreased exciton binding energies warrants optimization of synthesis procedures to experimentally prepare $Cs_2AgBi_{1-x}Fe_xBr_6$ with $x > 0.5$.

## 3. Conclusion

In this work we synthesized $Cs_2AgBi_{1-x}Fe_xBr_6$ solid solutions with different Bi:Fe ratios, up to $x = 0.30$, using a ball mill. Substitution of $Bi^{3+}$ with $Fe^{3+}$ leads to control over the visible light absorption of these materials between 2.1 and ~1 eV. Despite the improved visible light absorption, the photoredox activity is reduced upon introduction of $Fe^{3+}$. We attribute this to a lowering of the conduction band minimum as predicted by first-principles DFT. We show that introducing $Fe^{3+}$ leads to a decrease in the energy and restructuring of the CBM. Moreover, for high concentrations of $Fe^{3+}$ ($x > 0.5$) the bandgap is predicted to change from indirect to direct and the CBM becomes disperse due to hybridization between Ag 5s and Fe 3d conduction band states. The formation of such a highly delocalized CBM underlines the potential of these materials as visible light absorbers. Our dye decolorization experiments and DFT calculations further confirm the tunability of the CBM upon alloying with $Fe^{3+}$. The observed quenching of the PL of $Cs_2AgBi_{1-x}Fe_xBr_6$ suggests the presence of non-radiative sub-bandgap states. However, the high tunability of the bandgap, potential magnetic ordering effects, chemical stability and high absorptivity of organic dyes highlight the potential of these materials for photoconversion applications.

## 4. Experimental Section/Methods

*Chemicals.* Cesium bromide (CsBr), silver bromide (AgBr), bismuth bromide and 99.999% (metal basis) ($BiBr_3$) were bought from Alfa Aesar. Iron bromide 98% ($FeBr_3$) was bought from Sigma Aldrich/Merck. No further purification steps were taken. Anhydrous isopropanol and absolute ethanol were bought from Sigma Aldrich/Merck.

*Synthesis of $Cs_2AgBi_{1-x}Fe_xBr_6$ photocatalyst.* To synthesize $Cs_2AgBiBr_6$ we used a Retsch MM400 ball mill. To a 10 mL stainless steel jar we added 4 mmol CsBr, 2 mmol AgBr and 2 mmol $BiBr_3$ together with two 10 mm in diameter stainless steel beads. The $Cs_2AgBi_{1-x}Fe_xBr_6$ compositions were obtained by adjusting the $BiBr_3$:$FeBr_3$ ratio stoichiometrically, while keeping the total amount at 2 mmol. The photocatalyst powders were obtained after grinding for 90 minutes at 30 Hz. Unreacted $FeBr_3$ was removed by washing the products with absolute ethanol.

*Characterization.* Elemental analysis was done using Scanning Electron Microscopy Energy Dispersive X-ray spectroscopy (SEM–EDX). This was performed on a FEI Helios NanoLab G3 UC microscope equipped with an Oxford instruments silicon drift detector X-Max energy-dispersive spectroscope measuring at 15 keV and 0.1 nA. The samples were treated beforehand by sputtering a 10 nm Pt coating with a 108 Cressington manual sputter coater.

Powder X-ray diffraction (XRD) patterns were obtained using a Bruker AXs D8 Phaser powder X-ray diffractometer equipped with a Cu K$\alpha_{1,2}$ ($\lambda = 1.54184$ Å) radiation source operating at 40 kV and 40 mA. The patterns were recorded by measuring in Bragg–Brentano geometry at diffraction angles from 5° to 60° 2θ, with a step size of 0.02° and a response time of 0.5 s. The calculated diffractograms of $Cs_2AgBi_{1-x}Fe_xBr_6$ were determined using VESTA, considering Fm3m symmetry, $a = 11$ Å, for different Bi:Fe ratios.

Diffuse reflectance UV–vis spectra were obtained using a Perkin–Elmer lambda UV–Vis–NIR–lambda950S spectrometer equipped with an integrating sphere. The background was obtained by placing the sample holder filled with PTFE (400 μm grain size) at the sample position. The spectra were recorded from 1500 to 350 nm with a step size of 2 nm and an integration time of 0.4 s.

The photoluminescence spectra were recorded using a homebuilt optical setup consisting of a Nikon Ti-U inverted microscope body. A 405-nm laser (Picoquant D-C 405, with Picoquant PDL 800-D laser driver) was guided to the sample by a dichroic mirror (edge at 425 nm, Thorlabs DMLP425R) and focused by an oil-immersion objective (Nikon CFI Plan Apochromat Lambda 100×, NA 1.45) onto the sample. Emission was collected by the same objective and guided to a spectrometer (Andor Kymera 193i, 150 lines/mm reflective grating) connected to an electron-multiplying CCD detector (Andor iXon Ultra 888).

*Density Functional Theory calculations.* All density functional theory (DFT) calculations were performed with the Vienna-Ab-Initio Simulation Package (VASP).48 A plane-wave basis was used in combination with the projector augmented wave (PAW) method49 with the following atomic configuration: $5d^{10}6s^26p^3$ for Bi, $3p^63d^74s^1$ for Fe, $4s^24p^64d^{10}5s^1$ for Ag, $5s^25p^65s^1$ for Cs and $4s^24p^5$ for Br. Starting from the $Fm\bar{3}m$ space group in the conventional unit cell of $Cs_2AgBiBr_6$ with 4 Bi atoms, 4 Ag atoms, 8 Cs atoms and 24 Br atoms, we substituted $Bi^{3+}$ by $Fe^{3+}$, so that the ratio of Bi:Fe is 4:0, 3:1, 2:2, 1:3, 0:4 for the concentrations $x = 0$, $x = 0.25$, $x = 0.5$, $x = 0.75$ and $x = 1$ respectively. For each composition, the structure was relaxed using the SCAN functional40 allowing both cell parameters and atomic positions to vary. The plane-wave basis set was expanded using a cutoff energy of 600 eV and the Brillouin zone was sampled using a uniform un-shifted 4x4x4 $k$-point mesh. The irreducible points generated are 12, 12, 32, 12 and 10 for the concentrations $x = 0$, $x = 0.25$, $x = 0.5$, $x = 0.75$ and $x = 1$ respectively. The electronic structure of the alloys was studied using the PBE0 hybrid functional with spin orbit coupling (SOC), which slightly overestimates the bandgap of $Cs_2AgBiBr_6$ but describes the highly localized Fe 3d orbitals of the alloys well.

*Photocatalytic degradation of methylene blue.* 40 mg of photocatalyst was suspended in a 10 mg/L MB in ethanol solution. The mixture was subsequently stirred at 500 rpm in the dark for an hour to reach sorption equilibrium. After 60 minutes the suspension was illuminated with a blue (445 nm, FWHM = 30 nm) or green (550 nm, FWHM = 30nm) lamp. Aliquots were taken at 20-minutes time intervals to study the MB concentration over time. The MB concentration was determined by integrating the absorbance band between 600 and 700 nm recorded using an Agilent Technologies Cary 60 UV–Vis spectrometer. Before measuring the absorbance, the aliquots were diluted by a factor of 3 by adding anhydrous ethanol after which the diluted suspension was centrifuged at 1000 rpm for 5 minutes to remove the photocatalyst powder.

**Supporting Information**

Supporting Information is available from the Wiley Online Library or from the author.

**Acknowledgements**

E.M.H. acknowledges funding from the Dutch Research Council (NWO) under the grant number VI.Veni.192.034. H.J.J. and E.M.H. are further supported by the Advanced Research Center Chemical Building Blocks Consortium (ARC CBBC). L.A.M. received funding from the NWO under the grant number OCENW.XS22.2.039. The authors wants to thank dr. Freddy Rabouw for facilitating the equipment for the PL experiments and Tim Prins for the useful discussions about X-ray diffraction. K.F. and L.L. acknowledge funding from the NWO under grant number OCENW.M20.337 and computational resources provided by the Dutch national supercomputing center Snellius supported by the SURF cooperative.